\documentclass[preprint]{aastex}

\newcommand{\src}{4U~1916$-$053}
\newcommand{\xte}{{\it RXTE}}

\shortauthors{Galloway}
\shorttitle{Burst oscillations in \src}

\begin{document}


\title{Discovery of a 270~Hz X-Ray Burst Oscillation in the X-Ray Dipper \src}


\author{Duncan K. Galloway, Deepto Chakrabarty, Michael P. Muno, and
  Pavlin Savov}
\affil{\footnotesize Center for Space Research and Department of Physics,
  Massachusetts Institute of Technology, \\Cambridge, MA 02139}
\email{duncan,deepto,muno,pavlin@space.mit.edu}

\author{\ }
\affil{Accepted for {\sc The Astrophysical Journal Letters}, 2000 Dec 20}



\begin{abstract}
We report the discovery of a highly coherent oscillation in a type-I
X-ray burst observed from \src\ by the {\it Rossi X-ray Timing
Explorer}\/ (\xte\/).  The oscillation was most strongly detected
$\approx 1$~s after the burst onset at a frequency of 269.3~Hz, and it
increased in frequency over the following 4 seconds of the burst decay
to a maximum of $\simeq$272~Hz. The total measured drift of
$3.58\pm0.41$~Hz ($1\sigma$) represents the largest fractional change in
frequency ($1.32\pm0.15$~\%) yet observed in any burst oscillation.
If the asymptotic frequency of the oscillation is interpreted in terms of
a decoupled surface burning layer, the implied neutron star spin period is
around 3.7~ms.  However, the expansion of the burning layer required to
explain the frequency drift during the burst is around 80~m, substantially
larger than expected theoretically (assuming rigid rotation).  The
oscillation was not present in the persistent emission before the burst,
nor in the initial rise. When detected its amplitude was 6--12\% (RMS)
with a roughly sinusoidal profile. The burst containing the oscillation
showed no evidence for photospheric radius expansion, while at least 5 of
the other 9 bursts observed from the source by \xte\/ during 1996 and 1998
did.  No comparable oscillations were detected in the other bursts.  A
pair of kilohertz quasi-periodic oscillations (QPOs) has been previously
reported from this source with a mean separation of $348\pm 12$~Hz.  \src\
is the first example of a source where the burst oscillation frequency is
significantly smaller than the frequency separation of the kHz QPOs.
\end{abstract}




\section{Introduction}

Type I X-ray bursts are signatures of unstable nuclear burning on the
surface of accreting neutron stars (see \citealt*{lew95} and
\citealt{bil98a} for recent reviews).  Strong evidence for surface
brightness anisotropies during the bursts comes from observations of
highly coherent `burst oscillations' in 7 sources to date
(\citealt{stroh96}; see also \citealt{vdk00} and references therein).
These oscillations have been suggested to result from initially localized
nuclear burning, which spreads over the surface of the neutron star during
the early stages of the burst \cite[]{stroh96}.
As the burst evolves, the oscillations typically increase in frequency by
a few Hz; in X1658$-$298 the increase is $\approx5$~Hz, the greatest
increase seen so far \cite[]{wij00}.  This frequency drift may occur as a
consequence of the burning layer becoming decoupled from the star
\cite[]{stroh97,cb00}.  In several sources the oscillation approaches an
asymptotic frequency that may be the spin frequency $\nu_{\rm spin}$ of
the neutron star.  Although many observational characteristics of burst
oscillations seem to be consistent with this picture, a growing body of
observations points to a substantially more complex underlying mechanism
\cite[e.g.][]{mill99,stroh00}.  In this {\it Letter} we report the
discovery of a new burst oscillation in \src.

The low-mass X-ray binary (LMXB) \src\ ($l=31.4\arcdeg$, $b=-8.5\arcdeg$)
was discovered by the {\it Uhuru}\/ satellite in 1977 \cite[]{4ucat}.
{\it EXOSAT} observations revealed irregular X-ray dipping behavior with
a period of $\approx50$~min \cite[]{walter82,white82}, which optical
observations of the $m_V=21$ companion confirmed was approximately the
orbital period \cite[]{grin88}.
Analysis of \xte\/ observations during 1996 shows that the  source is a
member of the `atoll' class \cite[]{bloser00,boirin00b}.  Both high ($\ga
100$~Hz) and low frequency QPOs were detected in the power spectra of the
source as measured by \xte.  In particular, a pair of high-frequency QPOs
(`kilohertz QPOs') with a mean separation of $348\pm 12$~Hz was
simultaneously detected on five occasions \cite[]{boirin00b}.  On four of
these occasions, the observed separation was consistent with the mean
value, but on one occasion was only $290\pm 5$~Hz.  X-ray bursts
exhibiting photospheric radius expansion resulting from super-Eddington
luminosities indicate a source distance of 8.4--10.8~kpc \cite[]{smale88}.
The X-ray bursts observed by \xte\ during 1996 were searched for
high-frequency oscillations, but none were found to a $3\sigma$ limit of
3\% RMS amplitude \cite[]{boirin00b}. Here we present the results of a
more comprehensive timing study of X-ray bursts in \src\ observed by \xte,
including both the previously analyzed 1996 data and more recent
observations made during 1998.

\section{Observations and Analysis}

\src\ was observed using the \xte\/ Proportional Counter Array 
\cite[PCA;][]{xte96} throughout 1996 and during 1998 June--August.
These observations comprise 40 separate pointings with a total exposure
time of 323~ks. The data were screened to exclude earth occultations and
intervals of unstable pointing, with the maximum allowed offset
$0.01\arcdeg$.  A search of the {\tt Standard1} mode data (2--60~keV,
0.125~s time resolution, no energy resolution) revealed four type I bursts
during 1996 and six during 1998\footnote{An apparent weak burst on 1996
March 13 was on closer examination present in only one of the five
proportional counter units (PCUs). We attribute this to a detector
breakdown in PCU 3, the earliest example of which was previously thought
to occur several days later.} (see Table \ref{obs}). It is possible that
one or more of the observed bursts originated from a source other than
\src\ in the PCA field (FWHM $=1\arcdeg$).  The nearest
($\approx1\arcdeg$) known source from the Simbad catalog ({\url
http://simbad.u-strasbg.fr/Simbad}) is RXS~192242.1-051559, from which no
bursts have been reported.  We consider the likelihood of an unknown
bursting source within the PCA field of view at this relatively high
Galactic latitude remote.  As already noted by \cite{chou00}, the 1996
bursts are clustered in binary phase near the X-ray dip.  Interestingly,
the 1998 bursts do not show the same evidence for clustering, although the
Chou et al.  ephemeris still correctly predicts the X-ray dip times in the
1998 data.

The burst profiles typically exhibit a broad and somewhat flat-topped
peak, often followed by one or more narrow dips during the decay.
Uninterrupted PCA data from high-time-resolution modes (typically the {\tt
E\_125us\_64M\_0\_1s} mode with 122~$\mu$s time resolution and 64 channel
energy resolution) were available for 8 of the 10 bursts observed.  The
spectral characteristics of these bursts were measured by fitting absorbed
blackbody spectra in the 2.5--40~keV range to 0.25~s segments of data
covering the burst.  Spectral fitting was undertaken with {\sc xspec}
version 11 \cite[]{xspec}. A spectrum extracted from a short interval of
data immediately before the burst was used as the background.  Five of the
bursts showed an increase in the fitted blackbody radius coupled with a
decrease in the effective temperature during the first 10~s.  This
strongly suggests that photospheric radius expansion occurred during these
bursts. 
No evidence for radius expansion was found for the bursts on 1998 July 24,
August 1, and August 10.

\begin{figure}
\plotone{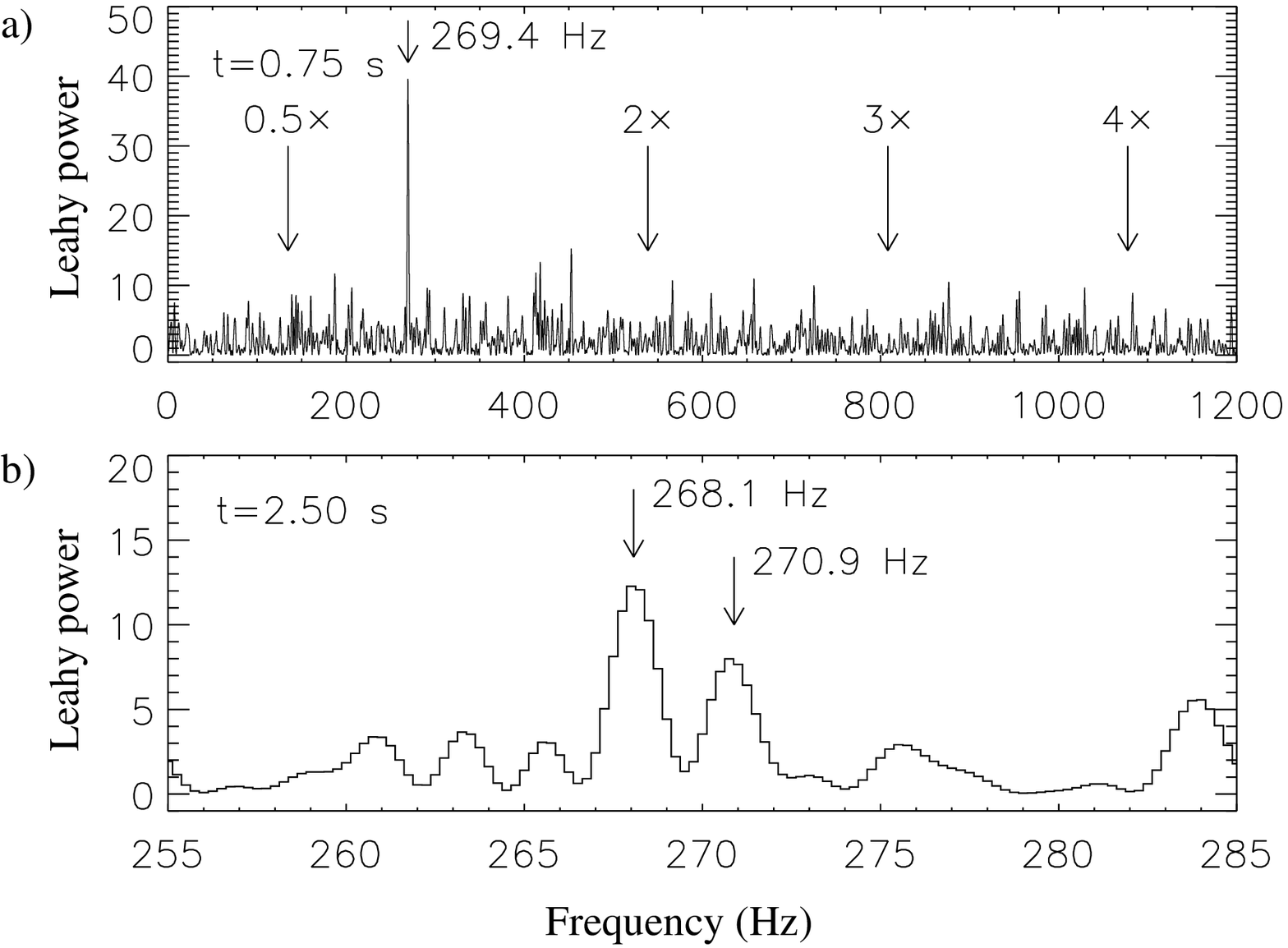}
 \figcaption[fig1.eps]{Leahy-normalized power spectrum from \xte\
observations of \src\ on 1998 August 1. The 8-times oversampled power
spectrum is calculated from 0.5~s segments of data, with
frequency resolution 0.25~Hz.
a) Power spectrum from 0.5~s segment beginning 0.75~s after the start of
the burst.  The largest peak is at a significance level of 99.99\% and
indicates a burst oscillation at 269.4~Hz. Expected positions of the $n=2$,3,4
harmonics and the first subharmonic are indicated by the arrows.
b) Power spectrum for the 0.5~s segment beginning 2.5~s after the start of
the burst. The less significant marked peak is close to the (single) peaks
observed in subsequent power spectra.
 \label{pwr} }
\end{figure}

\begin{figure}
 \epsscale{0.7}
 \plotone{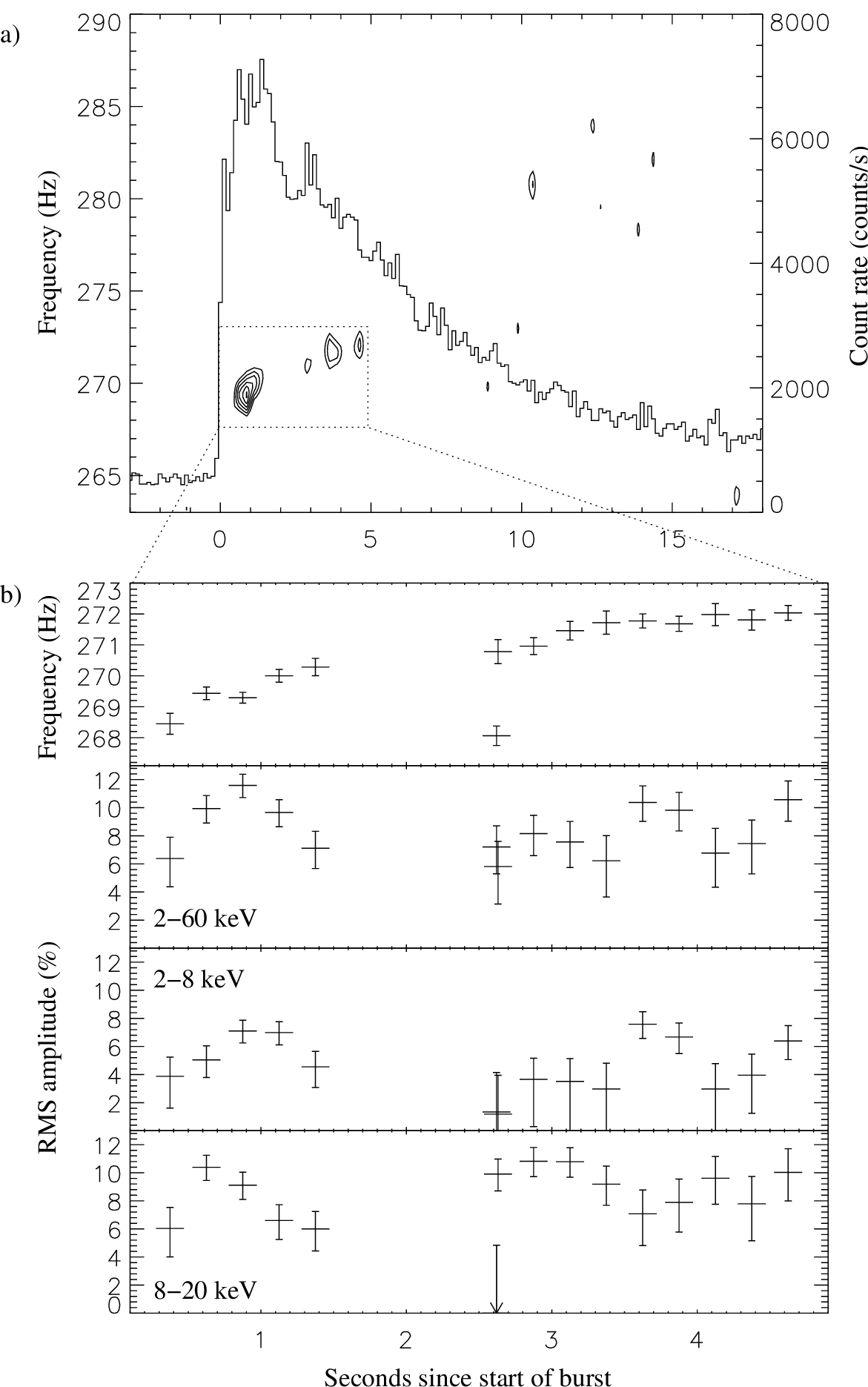}
 \figcaption[fig2.eps]{Time evolution of the oscillation in the 1998
 August 1 burst from \src. (a)~Dynamic power spectrum of the burst
 oscillation (contours) overplotted on the burst intensity profile
 (solid line).  The contour levels start at a Leahy-normalized power
 corresponding to 90\%-confidence and increase in steps of 5 in
 power. The highest contour level corresponds to a confidence level of
 $>99.99$\%.  (b) Time evolution of the oscillation frequency and RMS
 amplitudes. The top panel plots $\nu_{\rm \,burst}$ calculated from the
oversampled Fourier transforms. The three panels below give the RMS
amplitude in the full PCA range (2--60~keV) and the sub-ranges 2--8 and
8--20~keV respectively. Error bars represent the $1\sigma$ uncertainties.
\label{aug01} }
\end{figure}

Sub-intervals of the high--time-resolution data covering each of the
bursts were searched for oscillations in the 0--4000~Hz range.  We
computed $8\times$ oversampled Fourier power spectra for 0.5~s lengths of
data spaced at 0.25~s intervals and searched for statistically significant
pulsed signals.  A strong, highly significant
(chance probability for 2000 independent frequencies $5\times10^{-6}$,
equivalent to $4.6\sigma$) oscillation at $\nu_{\rm \,burst}\approx
270$~Hz was detected 0.75~s after the burst rise during the 1998 August 1
burst (Figure \ref{pwr}).  Weaker oscillations were measured at a
gradually increasing frequency over the next $\approx 5$~s, except for the
interval between 2 and 3~s following the burst rise when no significant
oscillation was detected in the 220--320~Hz range (Figure \ref{aug01}a).
When the oscillation reappeared 3~s after the burst rise, we initially
detected two peaks of approximately equal strength separated by
$\approx3$~Hz, although clearly only one was consistent with the
continuing evolution of the initial signal.  We note that the presence of
two closely separated peaks has also been observed in 4U~1636$-$536
\cite[]{mill00}.  The oscillation was not detected in the persistent
emission prior to the burst, nor during the burst rise itself.  The
frequency and RMS amplitude for the oscillation corresponding to each peak
was measured from the oversampled Fourier power spectra
\cite[e.g.][]{mn76}. The difference between the initial and final
frequencies at which the oscillation was detected ($268.45\pm0.34$ and
$272.03\pm0.24$~Hz respectively) was $3.58\pm0.41$~Hz ($1\sigma$ error;
Figure \ref{aug01}b).  The RMS amplitude for the full PCA energy range
(2--60~keV) peaked at $\approx$12\% around 1~s following the burst rise,
and then decreased within 0.5~s to unmeasurable levels.  When the
oscillation returned $\approx3$~s following the burst rise, the measured
amplitude was in the 6--11\% range.  The RMS amplitudes were also
calculated in the 2--8 and 8--20~keV energy ranges separately. The
amplitude was generally significantly greater in the 8--20~keV energy
band, at 6--12\%. However, at times the amplitude in the 2--8~keV band
reached that of the higher energy band, suggesting spectral variations in
the pulsed emission. We note that when two significant peaks close to
270~Hz were measured in the Fourier spectrum (at around 2.5~s following
the burst onset) only the higher frequency oscillation was present to a
significant level in the 8--20~keV energy band.  The pulse profile was
generally consistent with a sinusoid.

No significant persistent oscillations were observed in the other 9
bursts detected from the source in 1996 and 1998. A more sensitive search
was undertaken on those bursts by averaging power spectra over 1~s
segments of data within intervals of varying length covering the burst
peak, but this too resulted in no detections.  In several of the bursts
(1996 August 16 in particular), peaks representing detections at $>90$\%
confidence around 256 or 270~Hz were found in single power spectra as
early as 2~s prior to the burst or coincident with the burst rise.  These
peaks did not persist for more than 0.5~s, nor were they as significant
in power spectra of longer stretches of data. The most significant peak
($3\sigma$ equivalent for 4000 independent frequencies) was found in both
the 0.5 and 1~s power spectra around 4~s before the 1996 August 16 burst,
with frequency $278.01\pm0.07$~Hz ($1\sigma$) and estimated RMS amplitude
24\%. We include these results for completeness; without repeated
detections or greater significance their authenticity is questionable.

\section{Discussion}

We have conclusively detected a highly coherent $\simeq$270~Hz oscillation
during a type I X-ray burst from \src, bringing the total number of known
burst oscillation sources to eight.  If we interpret the frequency
evolution of the oscillation during the burst in terms of a decoupled
burning layer, then the maximum observed frequency implies a neutron star
spin period of $\approx 3.7$~ms.  However, the large change in frequency
during this burst ($3.58\pm0.41$~Hz or $1.32\pm0.15$~\%) would require an
$\approx80$~m expansion of the burning layer (assuming rigid rotation).
The maximum peak flux of all the bursts observed is a lower limit to the
Eddington flux $F_{\rm Edd}$ for the source, in which case the peak flux
for the 1998 August 1 burst (in which the burst oscillation was observed)
is at most $0.5 F_{\rm Edd}$. The maximum expansion predicted for a mixed
H/He burst at this flux level is only 20~m \cite[]{cb00}, different from
our measurement at the $7\sigma$ level.  That the binary period of the
source is shorter than 80~min indicates that the mass donor must be very
hydrogen-poor \cite[]{nrj86}, suggesting a predominantly He burning layer.
Interestingly, \cite{cb00} predict an even smaller expansion in that case.
The limited sample available suggests that burst oscillations in \src\ are
more likely to be present in atypically weak bursts which do not exhibit
photospheric radius expansion (see Table \ref{obs}).  The other two
sources with $\sim300$~Hz burst oscillations (4U~1728$-$34 and
4U~1702$-$429) also show this tendency, while the sources with higher
frequency oscillations behave in a distinctly different manner (Muno et
al. 2001, in preparation).



In the four burst sources from which both a kHz QPO pair and a burst
oscillation were previously known, the burst oscillation frequency
$\nu_{\rm \,burst}$ is comparable to (or else nearly twice) the peak
separation of the kHz QPOs $\Delta\nu$ (see Table~2).  Precise
measurements of $\Delta\nu$ have been made for two of these sources,
and these detailed measurements show clearly that
\begin{equation}
  \nu_{\rm \,burst} \gtrsim n\,\Delta\nu \mbox{\rm\ \ with $n=1$ or 2.} 
\end{equation}
\cite[]{mend98,mend99b}.
For the other two sources, one can only conclude that $\Delta\nu$ and
$\nu_{\rm \,burst}$ are equal within their uncertainties.  The
relationship between $\nu_{\rm \,burst}$ and $\Delta\nu$ given in (1)
has been interpreted in terms of a beat frequency model for the kHz QPOs 
(\citealt{stroh96}; \citealt*{mlp98}).  In the original sonic-point beat
frequency model of \cite{mlp98}, the upper kHz QPO appears at the
Keplerian frequency of the accretion flow's sonic point, and the lower kHz
QPO appears at the beat between this frequency and $\nu_s$.  In this
simple picture, we expect $\Delta\nu=\nu_{\rm spin}$.  More recent
calculations by \cite{lm00} have shown that the accreting material's
inward drift decreases the upper kHz QPO frequency below the sonic point's
Keplerian frequency and increases the lower kHz QPO frequency above the
beat frequency, resulting in $\Delta\nu\lesssim\nu_{\rm spin}$.  This is
consistent with the observed relation (1) if we have $\nu_{\rm
\,burst}\approx n\,\nu_{\rm spin}$ with $n=1$ or 2, as in the decoupled
burning layer model for the burst oscillations.  

Our study of \src\ has revealed the first example of a source where
$\nu_{\rm \,burst} < \Delta\nu$ to high significance ($\ga7\sigma$ for
the range of values of $\Delta\nu$ reported in \citealt{boirin00b}).  It
is unclear how to understand this result in terms of the sonic point
model, which would require the frequency of the upper kHz QPO to be higher
than the Keplerian orbital frequency in order to match our observation.
The numerical simulations presented by \cite{lm00} do not show this trend,
but such a possibility cannot be excluded a priori.  A different
explanation for the kHz QPOs involving relativistic precession of the
accretion disk has also been proposed.  Here, the upper kHz QPO appears at
the Keplerian frequency at some characteristic disk radius, and the lower
kHz QPO occurs at the periastron precession frequency at the same radius
\cite[]{sv98,pn00}.  In this picture, $\Delta\nu$ is approximately equal
to the radial epicyclic frequency and is not inherently related to
$\nu_{\rm spin}$, so that the striking observed similarity between
$\nu_{\rm \,burst}$ and $\Delta\nu$ (or $2\Delta\nu$) is not explicitly
addressed.  An additional model is therefore required to understand the
burst oscillations in conjunction with the relativistic precession models
for the kHz QPOs. 

The requirements for such a model have been discussed, although none
has yet been developed quantitatively.  If we assume that $\nu_{\rm
\,burst}\approx \nu_{\rm spin}$, then it must be that $\nu_{\rm spin}$ is
driven towards the maximum epicyclic frequency \cite[]{stella99} or else
that the kHz QPOs only reach detectable amplitudes when
$\Delta\nu\approx \nu_{\rm spin}$ or $\Delta\nu\approx \nu_{\rm spin}/2$ due to a resonance effect \cite[]{psa00}
Alternatively, $\nu_{\rm \,burst}$ might not be related to
$\nu_{\rm spin}$ at all, but might instead be related to an accretion
disk mode that occurs near the maximum epicyclic frequency at the
innermost stable orbit \cite[]{tlm98,pn00},
such as the $g$-modes predicted in the case
of accreting black holes \cite[see e.g.][]{kato90,nw91}.
Unfortunately, the absence of a detailed link between the burst
oscillations and the relativistic precession model makes it impossible
to constrain this model using our results. 

\acknowledgments We thank Ed Morgan and Dimitrios Psaltis for useful
discussions, and Josh Grindlay for making a source ephemeris available
prior to publication.  This work was supported in part by the NASA
Long Term Space Astrophysics program under grant NAG 5-9184.

\begin{deluxetable}{llcccc}
\tablecaption{X-Ray Bursts from 4U 1916$-$053\label{obs} }
\tablewidth{0pt}
\tablehead{
  & \colhead{Start time of burst}& \colhead{Peak flux\tablenotemark{a}}
  & \colhead{Radius} & \colhead{Oscillation} & \colhead{Binary}\\
 \colhead{Observation ID} & \colhead{(UTC)} &
 \colhead{($10^{-8}\ {\rm erg\,cm^{-2}\,s^{-1}}$)} & \colhead{expansion?} & 
 \colhead{frequency (Hz)} & \colhead{phase\tablenotemark{b}}}
\startdata
10109-01-04-00 & 1996 May 5 22:18:08  & $2.17\pm0.26$ & ? & \nodata & 0.02\\ 
10109-01-05-00 & 1996 Jun 1 18:28:44  & $2.90\pm0.23$ & ? & \nodata & 0.92\\ 
10109-01-07-00 & 1996 Aug 16 12:27:37 & $3.06\pm0.17$ & Y &  \nodata & 0.04\\ 
10109-01-09-00 & 1996 Oct 29 06:57:51 & $2.72\pm0.14$ & Y & \nodata & 0.02\\ 
30066-01-02-07 & 1998 Jul 23 05:30:52 & $3.26\pm0.18$ & Y & \nodata & 0.12\\ 
30066-01-02-08 & 1998 Jul 23 11:44:37 & $3.34\pm0.19$ & Y & \nodata & 0.59\\ 
30066-01-03-00 & 1998 Jul 24 17:03:15 & $2.23\pm0.24$ & N & \nodata & 0.76\\ 
30066-01-03-02 & 1998 Jul 26 17:05:16 & $3.16\pm0.17$ & Y & \nodata & 0.38\\ 
30066-01-03-03 & 1998 Aug 1 18:23:45  & $1.62\pm0.13$ & N & 269--272 & 0.71\\ 
30066-01-03-04 & 1998 Aug 10 11:40:53 & $3.03\pm0.15$ & N & \nodata & 0.79\\ 
\enddata
\tablenotetext{a}{Bolometric flux calculated according to the blackbody
spectral fit parameters}
\tablenotetext{b}{Phase with respect to 50 min X-ray dip period, with 
dip phase at 0.0. From ephemeris of \cite{chou00}.}
\end{deluxetable}

\begin{deluxetable}{lccl}
\tablecaption{Burst Oscillation Frequencies and kHz QPO Peak Separations}
\tablewidth{0pt}
\tablehead{
 \colhead{Source} & \colhead{$\nu_{\rm \,burst}$ (Hz)} & 
 \colhead{$\Delta\nu$ (Hz)} & \colhead{Refs.} }
\startdata
4U 1636$-$53 & 291, 582 & 251(4) & 1, 2, 3 \\
4U 1702$-$43 & 330      & 315(11)--344(7) & 4, 5 \\
4U 1728$-$34 & 363      & 279(12)--349(2) & 6, 7 \\
KS 1731$-$260 & 524     & 260(10) &  8, 9\\
4U 1916$-053$ & 270 & 290(5)--350(25)& 10, 11\\
\enddata
\tablerefs{
(1) \citealt{stroh98b};
(2) \citealt{mill99};
(3) \citealt{mend98};
(4) \citealt{sm99};
(5) \citealt{mss99};
(6) \citealt{stroh96};
(7) \citealt{mend99b};
(8) \citealt{smb97}
(9) \citealt{wv97};
(10) \citealt{boirin00b};
(11) this work.}
\end{deluxetable}



\end{document}